\documentclass[prd,twocolumn,nofootinbib,aps,tightenlines,preprintnumbers,notitlepage,longbibliography,superscriptaddress]{revtex4-1}

\usepackage{amssymb}
\usepackage{amsmath}
\usepackage{graphicx}
\usepackage{hyperref}
\usepackage{natbib}
\usepackage{color}

\begin{document}

\title{Strong Lensing of Gamma Ray Bursts as a Probe of Compact Dark Matter}

\author{Lingyuan Ji}
\author{Ely D. Kovetz}
\author{Marc Kamionkowski}
\affiliation{Department of Physics and Astronomy, Johns Hopkins University, 3400 N. Charles St., Baltimore, MD 21218, USA}

\begin{abstract}
Compact dark matter has been efficiently constrained in the $M \lesssim 10\, M_\odot$ mass range by null searches for microlensing of stars in nearby galaxies.  Here we propose to probe the mass range $M\gtrsim 10\, M_\odot$ by seeking echoes in gamma-ray-burst light curves induced by strong lensing.
We show that strong gravitational lensing of gamma ray bursts (GRBs) by massive compact halo objects (MACHOs) generates superimposed GRB images with a characteristic time delay of $\gtrsim\!1\,{\rm ms}$ for $M\!\gtrsim\!10\,M_\odot$. Using dedicated simulations to capture the relevant phenomenology of the GRB prompt emission, we calculate the signal-to-noise ratio required to detect GRB lensing events as a function of the flux ratio and time delay between the lensed images.
We then analyze existing data from the Fermi/GBM and Swift/BAT instruments to assess their constraining power on the compact dark matter fraction $f_{\rm DM}$. We find that this data is noise limited, and therefore localization-based masking of background photons is a key ingredient. Future observatories with better sensitivity will be able to probe down to the $f_{\rm DM}\gtrsim1\%$ level  across the $10\,M_\odot\lesssim M\lesssim 1000\,M_\odot$ mass range.
\end{abstract}


\maketitle

\section{Introduction}

While various independent experiments have garnered evidence for the existence of dark matter for nearly 50 years, what constitutes it remains a mystery~\cite{Bertone:2004pz}. An important class of candidates to make up the dark matter are massive compact halo objects (MACHOs)~\cite{Allsman:2000kg}. The MACHO scenario is well constrained at the low mass range ($\lesssim 10 M_\odot$) by the fact that we only observe a small optical depth of gravitational microlensing for stars in nearby galaxies~\cite{Tisserand:2006zx, Wyrzykowski:2011tr}. Meanwhile, the higher mass range ($\gtrsim 100 M_\odot$) is constrained by various dynamical considerations (such as perturbing effects from MACHOs on Galactic wide binaries \cite{Quinn:2009zg, Yoo:2003fr, Monroy-Rodriguez:2014ula} or the compactness of star clusters and ultra-faint dwarf galaxies~\cite{Brandt:2016aco}), and by lack of radiation as a result of accretion from the CMB (if the compact dark matter are primordial black holes)~\cite{Ali-Haimoud:2016mbv, Horowitz:2016lib, Blum:2016cjs, Ricotti:2007au}.

The intermediate mass range, $10M_\odot \lesssim M_L \lesssim 100M_\odot$, has received a lot of attention following the recent LIGO detection of several merging black hole binaries~\cite{Abbott:2016blz} and the suggestion that these may be primordial black holes (PBHs)~\cite{Zeldovich1966, Hawking1971} which could comprise a significant fraction of the total dark matter in the Universe~\cite{Bird:2016dcv, Clesse:2016vqa, Sasaki:2016jop, Kashlinsky:2016sdv}. This mass range remains less robustly constrained~\cite{Sasaki:2018dmp}, and novel approaches to probing it are well motivated~\cite{Kovetz:2017rvv}.

In this paper we propose to seek MACHOs in the $\gtrsim 10\, M_\odot$ mass range by searching gamma-ray-burst (GRB) light curves for signs of microlensing induced echoes.
GRBs are extremely energetic explosions originating in distant galaxies that lead to rapid emission of high-energy gamma rays~\cite{Meszaros:2006rc, Peer:2015eek, Gehrels:2012kp}. Several thousands of GRBs have been detected to date by dedicated instruments such as BATSE, FERMI GBM and SWIFT BAT. The prompt emissions of observed GRBs last from $\sim\! 10\,{\rm ms}$ to $\sim\! {\rm hours}$ and the source redshifts are $\mathcal{O}(z=1)$.

Strong lensing by a massive compact lens will generate two images of the original GRB \cite{Kochanek:2004ua, Petters2001}. The angular separation between the images is too small to be probed by current observations \cite{Meegan:2009qu, Gehrels:2004aa, Paciesas2004}. But the arrival times of the two images differ by roughly the Schwarzchild crossing time, roughly $0.3\,(M_L/10\, M_\odot)$~milliseconds for a lens of mass $M_L$. Here we build upon Ref.~\cite{Munoz:2016tmg}, wherein it was suggested that these microlensing-induced echoes can be sought with fast radio bursts (FRBs), to explore the possibility to seek these echoes with GRBs. Unlike Fast Radio Bursts (FRBs)~\cite{Munoz:2016tmg}, whose intrinsic time width is short ($\sim1\,{\rm ms}$), the images of strongly lensed GRBs will not appear as separate consecutive bursts. Rather, the fluxes from the two images will be added up to give the total light curve measured by a GRB observatory. However, lensed GRBs with time delays longer than the minimum variability timescale (MVT) of the intrinsic burst signal could be resolved by the correlation between the displaced features in the overlaid first and second images, which will generate a peak in autocorrelation of the total light curve. As we show below, with $\text{MVT} \sim 1\,{\rm ms}$~\cite{MacLachlan:2012cd}, it is possible to probe MACHO lens masses $M_L\gtrsim10 M_\odot$.

Although detecting repeating FRBs is simpler due to their cleaner temporal shape, GRBs are detectable out to higher redshifts. As a result, for the same MACHO mass $M_L$ and fraction $f_\text{DM}$ of the total dark matter that is made up of MACHOs, the integrated lensing optical depth becomes larger, which increases the sensitivity of our probe. Furthermore, a higher lens redshift will yield a longer time delay, which extends the mass range we can probe at the lower end, given the same MVT.

The idea to seek lensing-induced GRB echoes traces back to Ref.~\cite{Mao:1993}, and there have been searches for GRB repeaters \cite{Quashnock:1993vp,Tegmark:1995ee,Hakkila:1997sn,Gorosabel:1998pm}. Refs.~\cite{Nemiroff:1994uj,Ougolnikov:2001xb,Nemiroff:2001bp,Veres:2009ji,Hirose:2006vj} studied repeated bursts induced by lensing. Ref.~\cite{Veres:2009ji} looked for similarities in light-curve shapes between different bursts.  Ref.~\cite{Hirose:2006vj} used a similar auto-correlation method as we do here to seek echoes within individual light curves. We extend upon this method by performing new simulation tests, including updated datasets, and focusing more on the push to shorter-timescale echoes, as we search for lensing induced by dark-matter lenses rather than Pop III stars.

In Section \ref{sec:theory}, we review the theory of strong gravitational lensing by a point mass lens and the calculation of the lensing optical depth. In Section \ref{sec:method}, we introduce an algorithm to detect strongly lensed GRB. In Section \ref{sec:data}, we reduce the Fermi Gamma-ray Burst Monitor (Fermi/GBM) and Swift Burst Alert Telescope (Swift/BAT) TTE data and show how to determine the detection cross Section from the data, based on a comparison with dedicated simulations we develop for this purpose. In Section \ref{sec:results}, we show the best constraints we can reach for various cases considered. We discuss future improvements in \ref{sec:discussion} and conclude in Section \ref{sec:conclusions}.

\section{Theory}

\label{sec:theory}

A MACHO of mass $M_L$ can be modeled as a point mass lens \cite{Kochanek:2004ua, Petters2001} with an angular Einstein radius
\begin{equation}
  \theta_{E}=2\sqrt{\frac{GM_{L}}{c^{2}}\frac{D_{LS}}{D_{S}D_{L}}},
\end{equation}
where $D_S$, $D_L$ and $D_{LS}$ are the angular diameter distances from the observer to the source, to the lens and between the source and  lens, respectively.
The two images produced by the gravitational lensing lie at angular positions
\begin{equation}
  \theta_\pm = \frac{\beta \pm \sqrt{\beta^2 + 4\theta_E^2}}{2},
\end{equation}
where $\beta$ is the angular impact parameter. The time delay between these two images can be shown to be~\cite{Munoz:2016tmg}
\begin{equation}
\label{eqn:Delta-t}
  \Delta t=\frac{4GM_{L}}{c^{3}}(1+z_{L})\left[\frac{y}{2}\sqrt{y^{2}+4}+\textnormal{ln}\left(\frac{\sqrt{y^{2}+4}+y}{\sqrt{y^{2}+4}-y}\right)\right],
\end{equation}
where $y\equiv\beta/\theta_E$ is the impact parameter normalized by the Einstein radius and $z_L$ is the lens redshift. The difference in luminosity between these two images can be characterized by a flux ratio $R$ defined as
\begin{equation}
\label{eqn:R}
  R\equiv \left|\frac{\mu_{+}}{\mu_{-}}\right|=\frac{(y^{2}+2)+y\sqrt{y^{2}+4}}{(y^{2}+2)-y\sqrt{y^{2}+4}},
\end{equation}
where $\mu_+$ and $\mu_-$ are the magnifications of the two images, respectively. Note that for a given lens mass $M_L$ and lens redshift $z_L$, both $R$ and $\Delta t$ as a function of $y$ are monotonic.

If the lensing-induced echo is to be detectable, the impact parameter must satisfy two conditions. First, there exists a minimum detectable time delay $\overline{\Delta t}$, where for any $\Delta t<\overline{\Delta t}$ the time delay between the images is too short to be resolved by autocorrelation method.  Secondly, there exists a maximum detectable magnification ratio $\overline{R}$, where for any $R>\overline{R}$ the second image is too weak to be  detected. The two factors determine an effective cross section,
\begin{align}
\begin{split}
	  &\sigma (M_L,z_L, z_S;\overline{R},\overline{\Delta t})\\
	  &=\pi \theta_E^2 \mathrm{Ramp}[y^2_\mathrm{max}(\overline{R})-y^2_\mathrm{min}(M_L,z_L,\overline{\Delta t})],
\end{split}
\end{align}
for a single lens given by an the annulus between the maximum and minimum impact parameters. Here $\mathrm{Ramp}(x)=(x+|x|)/2$ is the ramp function. Note that the implicit $z_S$ dependence  emerges through the angular diameter distances $D_{LS},D_S$  in $\theta_E$~\cite{Hogg:1999ad}.

To find the optical depth of a GRB at redshift $z_S$ to be lensed by a MACHO on its way to Earth, in the optically-thin regime, we need to integrate over the cross sections. In our model we take all MACHOs to be of the same mass and uniformly distributed in the Universe. This gives
\begin{align}
\begin{split}
  \tau(z_S,M_L,f_\mathrm{DM};\overline R, \overline{\Delta t})=&\int_0^{z_S} \mathrm{d}\chi(z_L)\ (1+z_L)^2n_L\\
  &\times \sigma (M_L,z_L, z_S;\overline{R},\overline{\Delta t})
  \end{split}
\end{align}
where $\chi(z)$ is the comoving distance as a function of redshift $z$,  $\mathrm{d}\chi(z_L)=c\ \mathrm{d}z_L/H(z_L)$, where $H(z_L)$ is the Hubble parameter at redshift $z_L$, and $n_L$ is the MACHO comoving number density (which can be related to $f_\mathrm{DM}$ by $f_\mathrm{DM}=8\pi G n_L M_L c^2/3H^2_0\Omega_\mathrm{d}$). Using the GRB redshift distribution $N(z_S)$ \cite{Coward:2012ay, Gruber:2014iza}, the total optical depth is
\begin{align}
\begin{split}
	&\tau_\text{tot}(M_L, f_\text{DM}; \overline R, \overline{\Delta t}) \\
	&=\int N(z_S)dz_S\ \tau(z_S,M_L,f_\mathrm{DM};\overline R, \overline{\Delta t}).
\end{split}
\end{align}

\begin{figure}
  \includegraphics[width=0.5\textwidth]{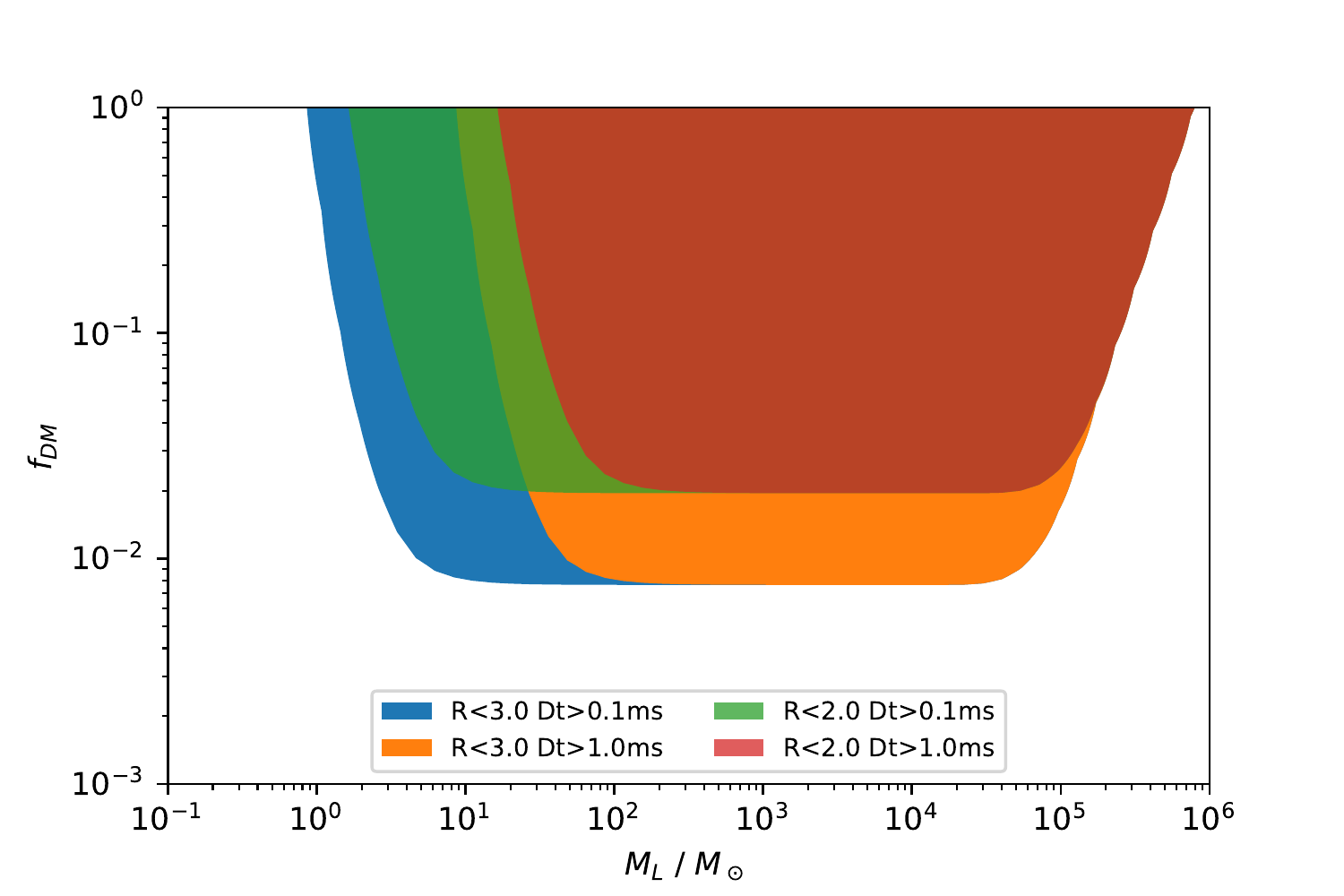}
  \caption{Fraction $f_\mathrm{DM}$ of dark matter than can be probed for different lens masses $M_L$ in a dataset of size $N\sim 2000$. Each patch is calculated with different minimum detectable time delay cutoff $\overline{\Delta t}$ and maximum detectable magnification ratio cutoff $\overline{R}$. In each case, the sensitivity is reduced at the low mass end because the generated time delay is too small to resolve; at the high mass range, the sensitivity is reduced because, practically, the detectable time delay will also have an upper bound due to the maximum correlation time probed. Here we take it to be $3\ \mathrm{secs}$.}
 \label{fig:constraints}
\end{figure}

If in a catalog containing $N$ GRBs, a certain algorithm can pick out no more than $N_*$ events as lensed, up to a specific cross section criterion $(\overline R, \overline{\Delta t})$, then $\tau_\text{tot}<N_*/N$ can be put as a constraint. As an example, we plot in Fig.~\ref{fig:constraints} the constraints to the $f_{\textrm{DM}}$-$M_L$ parameter space that would arise from a null search in a dataset of size $N\sim 2000$. Here some representative values of $\overline{\Delta t}$ and $\overline R$ are chosen. Comparing to the FRB case \cite{Munoz:2016tmg}, where the integrated optical depth is $\tau_{\rm tot}\sim 2\%$ (for $\overline{R}=5.0$ and $f_\text{DM}=1$) within the sensitive mass range, for GRBs it increases to $\tau_{\rm tot}\sim 15\%$ under the same conditions (and reduces to $\tau_{\rm tot}\sim 6\%$ for $\overline R=3.0$).

 In principle, the limitation on $\Delta t$ comes from the MVT of the light curve, while the limitation on $R$ is mainly related to the light curve's signal-to-noise ratio (SNR). The specific algorithm one uses to detect the lensing feature in the light curves will affect these two limitations as well.

\section{Method}
\label{sec:method}

\subsection{Light Curve Autocorrelation}
\label{sec:lcac}

In the case of FRBs~\cite{Munoz:2016tmg}---which are intrinsically very short ($\sim1\,{\rm ms}$)---a lensed burst would appear as two distinct peaks, whose time delay can then be read manifestly, provided that the flux ratio is not too small.
GRBs, however, are a different story, as their duration can range from milliseconds to minutes (and longer). The light curve of a a strongly lensed GRB will have an echo superimposed on the light curve. To detect this, we define the autocorrelation function,
\begin{equation}
  C(\delta t)\equiv\frac{\int dt\,I(t)I(t-\delta t)}{\sqrt{\int dt\,I^{2}(t)}\sqrt{\int dt\,I^{2}(t-\delta t)}},
\end{equation}
 of a light curve $I(t)$~\cite{Hirose:2006vj, Bracewell1986}. This is the normalized cross correlation  between a function and its shifted self. Note that $C(0)=1$. For generic non-autocorrelated functions, $C(\delta t\neq 0)<1$, and thus $\delta t\!=\!0$ serves as the only major spike in the  function $C(\delta t)$.

The effectiveness of  the autocorrelation analysis on strongly lensed data can be shown as follows. Let $I(t)$ be the intrinsic light curve of the GRB event and $C(\delta t)$ its autocorrelation. In a specific lensing configuration, Eqs.~\eqref{eqn:Delta-t} and \eqref{eqn:R} assign a time delay $\Delta t$ and a flux ratio $R$ to the second image, yielding a lensed light curve
\begin{equation}
\label{eq:lensed-light-curve}
  I(t;\Delta t,R)\propto \frac{R}{R+1}I(t)+\frac{1}{R+1}I(t-\Delta t).
\end{equation}
It is easy to see that the autocorrelation of the lensed light curve $I(t;\Delta t,R)$ is then given by
\begin{align}
\begin{split}
  \label{eq:lensed-autocorrelation}
  &C(\delta t;\Delta t,R)\\
  =&\frac{(R^{2}+1)C(\delta t)+R[C(\delta t+\Delta t)+C(\delta t-\Delta t)]}{(R^{2}+1)+2RC(\Delta t)}.
\end{split}
\end{align}
According to this formula, the lensed signal will exhibit spikes at $\delta t=-\Delta t,0,+\Delta t$ with an amplitude ratio $R/(R^2+1):1:R/(R^2+1)$.

\subsection{Detection Algorithm}
\label{sec:classifier}

\begin{figure}
	\includegraphics[width=0.5\textwidth]{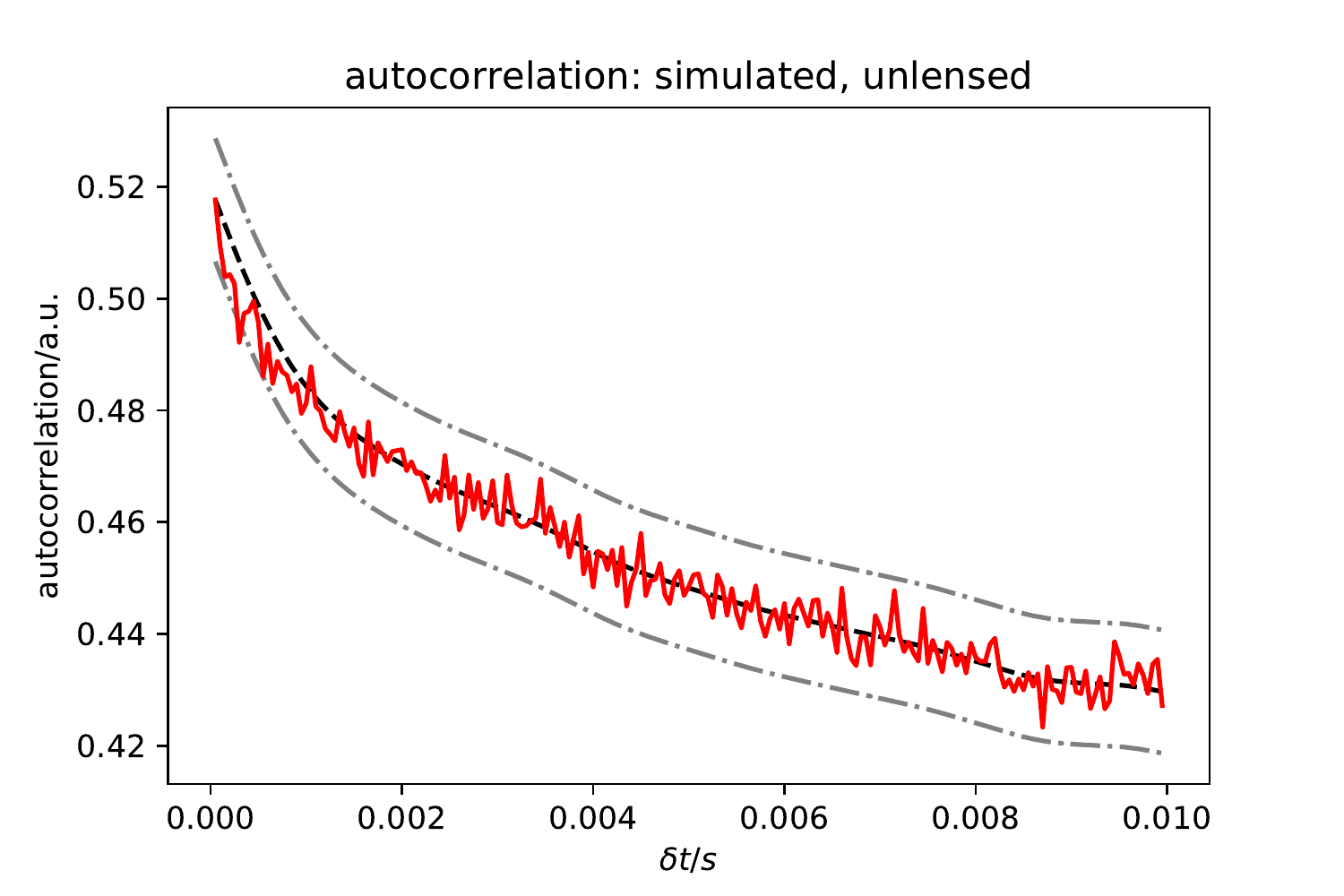}
	\includegraphics[width=0.5\textwidth]{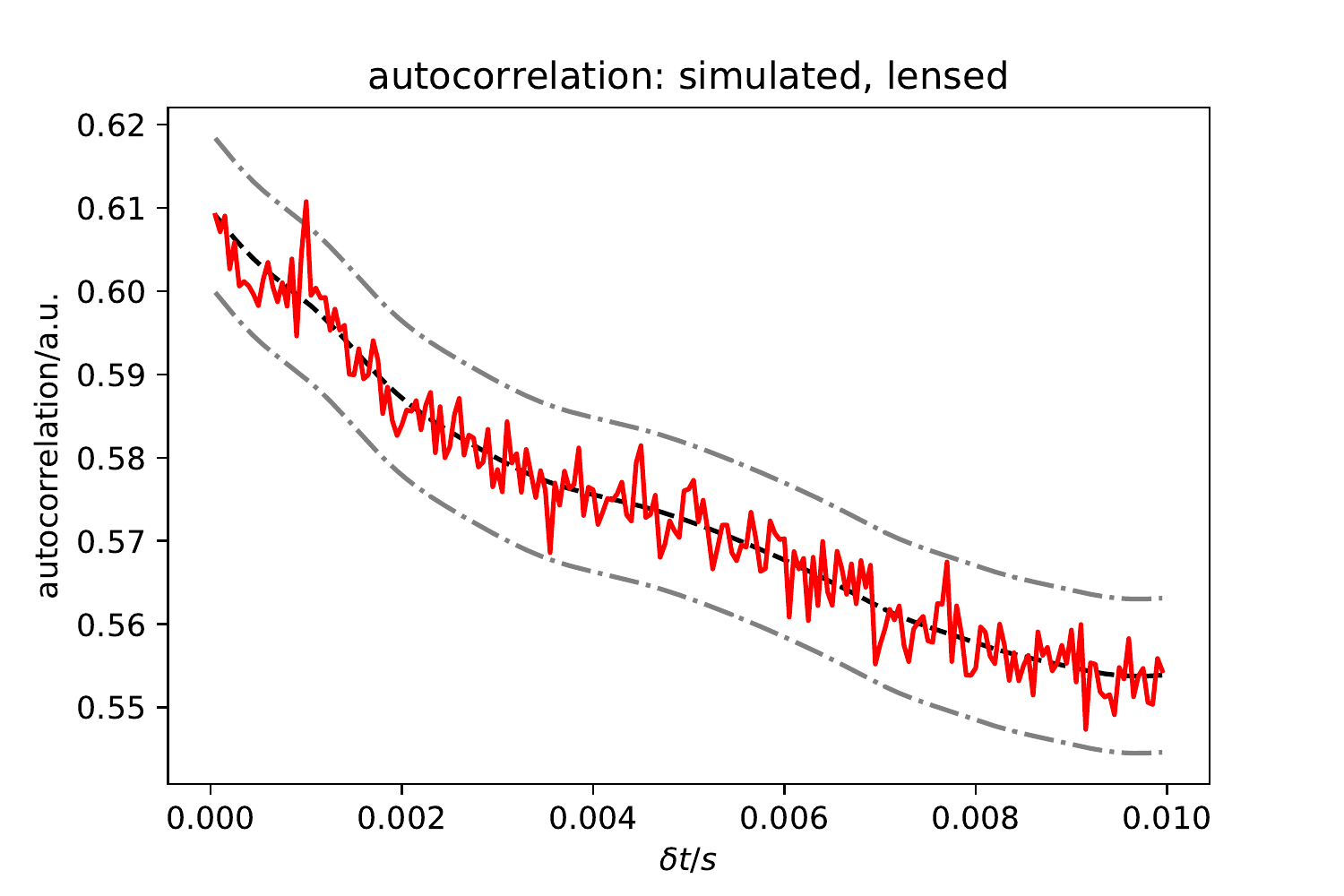}
	\caption{Red line represents the light curve autocorrelation. Black dashed line is the Gaussian smoothing of autocorrelation. Gray dotted dashed lines are $\pm 3 \sigma$. In the upper panel, no exceedance larger than $3\sigma$ is found, while in the lower panel there is an exceedance at $\delta t = 1\mathrm{ms}$. It can be seen that this algorithm can effectively detect lensing events at a certain level. More discussion about the sensitivity of the algorithm can be found in Section~\ref{sec:data} and how to simulate the light curve is elaborated in Section~\ref{sec:data}.}
	\label{fig:sample-ac}
\end{figure}

Real data, however, result in a spiky structure even for $C(\delta t)$, which means $\delta t=-\Delta t,0,+\Delta t$ are not the only spikes in $C(\delta t;\Delta t,R)$. To distinguish lensing-induced spikes from noise, we define the sigma parameter
\begin{equation}
  \sigma=\sqrt{\frac{\sum_{\delta t\in D}[C(\delta t)-G(\Delta t)]^{2}}{|D|}},
\end{equation}
where $C(\delta t)$ is the autocorrelation one wants to check (can be lensed or not), $G(\delta t)$ is a Gaussian smoothing of $C(\delta t)$ and $D$ is the time grid over which the calculation is performed. The sigma parameter quantifies the overall spikiness in $C(\delta t)$ compared to a smooth template $G(\delta t)$. If we identify a $\delta t$ for which
$  |G(\Delta t)-C(\delta t)|$ is larger than some threshold (e.g.\ $3\sigma$ or $4 \sigma$), we can claim evidence for a spike due to lensing. In Fig.~\ref{fig:sample-ac} we present the autocorrelations of a simulated unlensed/lensed GRB light curve. The simulation method is described in Section~\ref{sec:data}.

\subsection{Noise and Minimum Variability Timescales}

The two major reasons responsible for the limitations $\Delta t>\overline{\Delta t}$ and $R<\overline R$ in the gravitational lensing model are: (1) the amount of noise in the observed light curve; and (2) the minimum variability timescale. Eq.~\eqref{eq:lensed-autocorrelation}, the formulation of the autocorrelation of the lensed light curve, may help to show this. On the one hand, for the peaks at $\Delta t$ to be separable from the peak at $0$, the full width at half maximum (FWHM) should be smaller than $\Delta t$. However, the FWHM of an autocorrelation is approximately the signal's MVT. So we can take $\overline{\Delta t}\sim \text{MVT}$. On the other hand, the amplitude $R/(R^2+1)$ of the peak at $\Delta t$ has to compete with the amplitude of random peaks due to noise. This eventually sets $\overline R$.

It is worth stressing that there is a conceptual difference between this work and Ref.~\cite{Munoz:2016tmg}, which addressed strong lensing of FRBs by compact dark matter objects.
There, the minimum detectable time delay was set by the temporal width of the FRB, which is typically $\mathcal{O}(1)\,\mathrm{ms}$. Here, the temporal width of GRB is $T_{90}\sim \mathrm{seconds}$, which is much larger than the typical time delay for strong lensing by point source of $\gtrsim10\,M_\odot$. The relevant question for determining the detectability of a lensed GRB through autocorrelation is what is the smallest timescale that still contains variability features that can autocorrelate. This shifts the limitation on the time delay to the MVT. %

\section{Application}
\label{sec:data}

\subsection{Overview of GRB observatories}

\begin{figure}
	\includegraphics[width=0.5\textwidth]{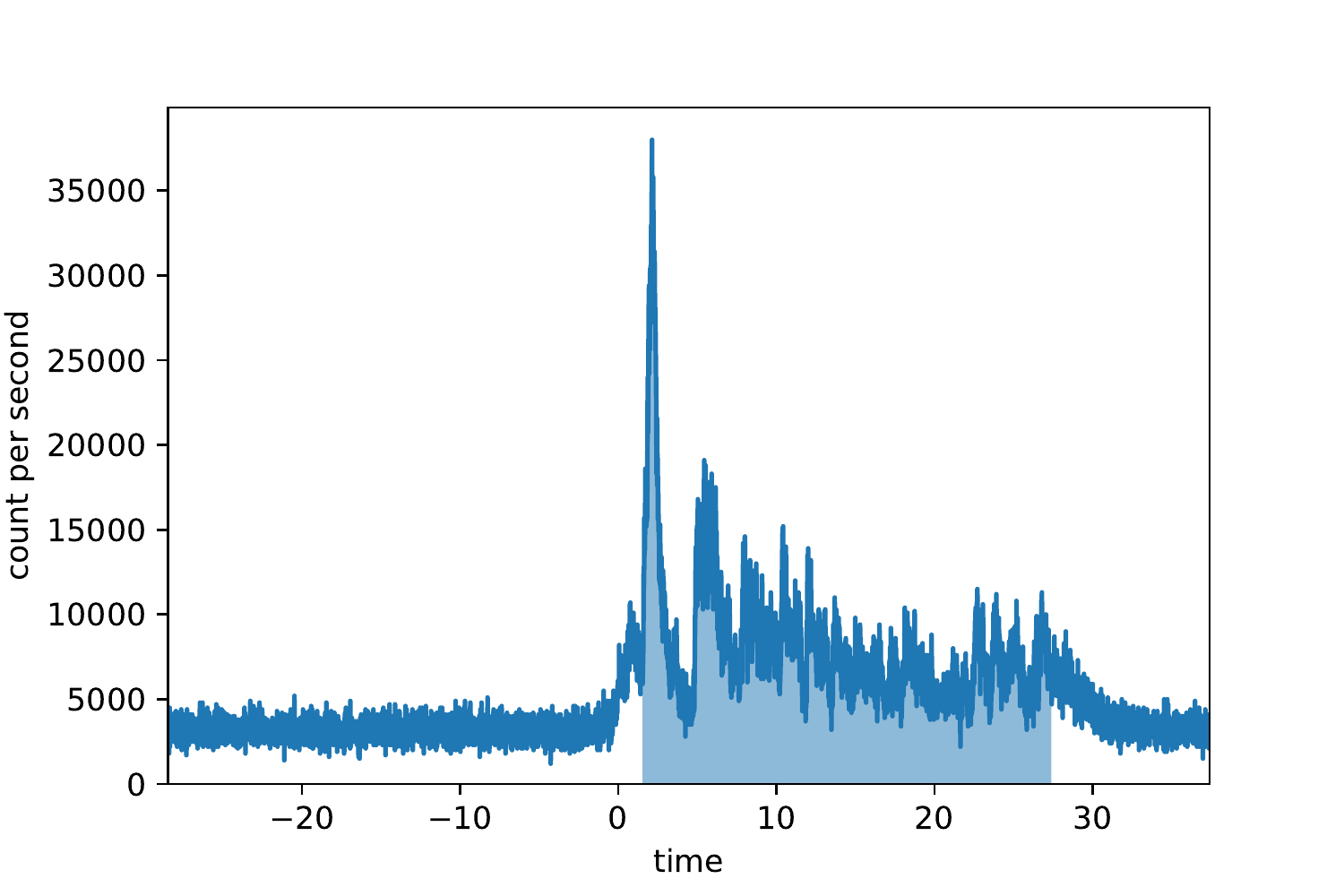}
	\caption{\texttt{bn161218356} light curve (from Fermi/GBM).
	The trigger time is set to be the origin, while $T_{90}$ range is shaded. The steady noise between $T_{90, \mathrm{start}}-20\mathrm s$ and $T_{90, \mathrm{start}}-10\mathrm s$ represents the noise. The $T_{90}$ range represents the signal.}\label{fig:sample-lc}
\end{figure}

From the previous Section, we see that the gravitational lensing time delay generated by $M_L\gtrsim10\,M_\odot$ lenses requires a time resolution better than $\sim\!1\, \mathrm{ms}$. Fermi/GBM and Swift/BAT properly suit our purpose \cite{Gruber:2014iza, Meegan:2009qu}. If the experiment's trigger condition is reached, all detectors' TTE data will be recorded for the following several hundreds seconds, and a mask will be generated for triggered detectors. We use the this mask to sum up all the photons for the triggered detectors to get the array $\{t_i\}_{i=1}^n$. The data is then binned to generate a light curve. We use a subset of the Fermi/GBM catalog\footnote{{\tt https://heasarc.gsfc.nasa.gov/FTP/fermi/data/gbm/bursts/}} with $N\!=\!2274$ GRBs ($N\!=\!1230$ for Swift/BAT\footnote{{\tt https://swift.gsfc.nasa.gov/results/batgrbcat/}}). An example light curve (trigger \# \texttt{bn161218356}) is shown in Fig.~\ref{fig:sample-lc}.

\subsection{Lensing Simulation}
\label{sec:lensing-experiment}

As presented in Section~\ref{sec:lcac}, a light curve model with the correct amount of noise added in is needed to obtain $\overline R$. Also, the MVT has to be incorporated in the model for us to verify the choice $\overline{\Delta t}=\text{MVT}$. Here we employ a model where the observed lensed light curve is given by
\begin{align}\label{eq:model}
\begin{split}
		&I(t;\Delta t, R; H, \text{MVT}, A)\\&=\frac{R}{1+R}B'_H(t)+\frac{1}{1+R}B'_H(t-\Delta t)+ A n(t),
\end{split}
\end{align}
where $B_H(t)$ is one realization of a fractional Brownian motion (fBm) \cite{Mandelbrot1968} with Hurst index $H\in(0,1)$, $B'_H(t)$ is $B_H(t)$ Gaussian filtered with radius $\text{MVT}/2\pi$, $n(t)$ is a unit Gaussian noise (GN) and $A$ quantifies the amplitude of noise in the measured light curve. We use fBm  because it contains features on all timescales, which enables us to cut off the power spectrum at a chosen high frequency to simulate the MVT behavior. In addition, the power spectrum of fBm scales as $1/f^{2H+1}$ \cite{Flandrin1989}, which resembles observed GRB light curves at low frequencies (Fig.~\ref{fig:sample-fft}). The value of $H$ will be determined below using the data. The term $B'_H(t)$ corresponds to the image that arrives first and the term $B'_H(t-\Delta t)$ to the delayed image. The coefficients in front are the amplitudes of each image, chosen to produce the magnification ratio $R$ while also maintaining the total signal strength the same as $B'_H(t)$.

In each lensing simulation, we fix the values of the parameters $\{H, \text{MVT}, A\}$ and sample the parameter space $(\Delta t, R)$. For each lensed light curve generated, the detection algorithm is applied to determine the significance of the lensing effect. A $3\sigma$ threshold is then used to set $\overline{\Delta t}$ and $\overline R$. To eliminate the bias from the specific profile of each specific light curve realization, the significance is averaged among an adequate amount of light curve realizations with the same set of chosen parameters.

To quantify the constraining power of Fermi/GBM or Swift/BAT, the parameters in our model, Eq.~\eqref{eq:model}, need to be inferred from the data. We do this next.

\subsection{Results}
\label{sec:results}

\begin{figure}
	\includegraphics[width=0.5\textwidth]{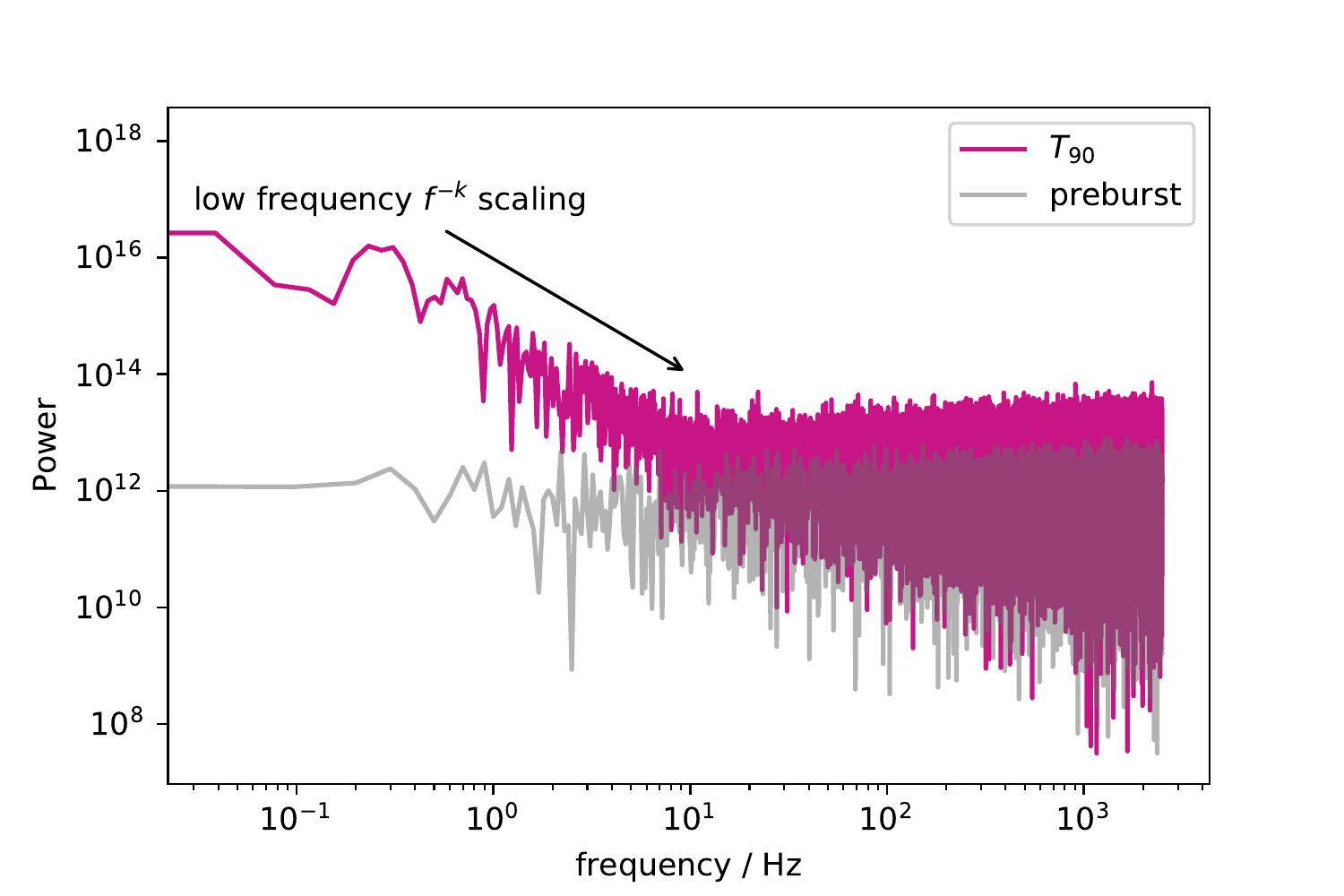}
	\caption{The power spectrum of \texttt{bn161218356}. For the $T_{90}$ range, a scaling feature $1/f^k$ can be seen at low frequency. In the pre-burst range, the power spectrum is scale independent, which is typical for gaussian noise.}\label{fig:sample-fft}
\end{figure}

\begin{figure}
	\includegraphics[width=0.5\textwidth]{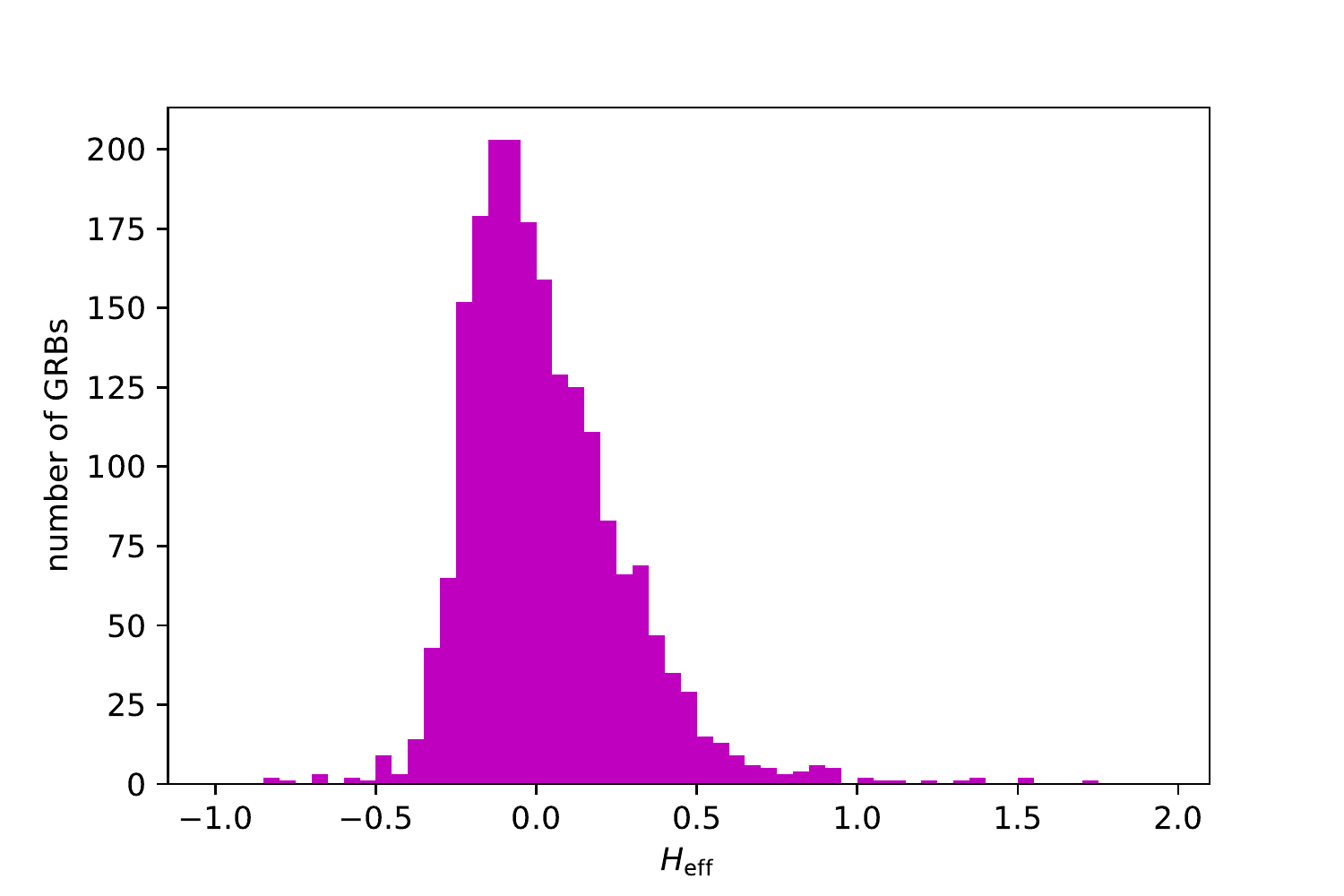}
	\caption{Effective Hurst index $H_\mathrm{eff}$ distribution for light curves in the Fermi/GBM catalog. The median of this distribution is approximately 0. Given that the allowed range for the Hurst index is $H\in(0,1)$, we choose $H=0.1$ in our simulations.}\label{fig:h-eff}
\end{figure}

We first use the data to infer the suitable value of  $H$. A fBm with Hurst index $H$ will have a power spectrum scaling as $P\propto 1/f^{2H+1}$. The same scaling behavior exists in the Fermi/GBM and the Swift/BAT data; see for instance Fig.~\ref{fig:sample-fft}. We therefore fit the low frequency ($0-2\ \mathrm{Hz}$) scaling part  of the power spectrum using the template $P\propto 1/f^k$, and then set $H_\mathrm{eff}\equiv(k-1)/2$. The distribution of $H_\mathrm{eff}$ for the Fermi/GBM catalog is shown in Fig.~\ref{fig:h-eff}. We see that the median value is $H_\mathrm{eff}\approx 0.0$. Since the allowed range for $H$ is $(0,1)$, we fix $H=0.1$.

The noise amplitude $A$ is also related to the power spectrum. While the light curve signal has the scaling property above, Gaussian noise has equal power at all timescales. Therefore, the noise power will always be dominant on the  smallest time scales, as shown in Fig.~\ref{fig:sample-fft}. It is useful to define the ratio
\begin{equation}
	r=\frac{P_\mathrm{median}(T_{90}\text{ range, 0-2 Hz})}{P_\mathrm{median}(\text{pre-burst, 10-30 Hz})}
\end{equation}
of these two powers. On a specific grid the lensing simulation is performed, $A$ can be expressed as a function of $r$. In our case, the simulation length is $1\,\mathrm s$ with $t_\mathrm{bin}\!=\!50\,\mu \mathrm s$. Numerical experiments show that a good approximation to take is
\begin{equation}\label{eq:r-A-relation}
	r\sim 1+\frac{50\mathrm{ms}}{t_\mathrm{bin}}\frac{1}{A^2}.
\end{equation}

The distributions of power ratio $r$ in the Fermi/GBM and Swift/BAT datasets are shown in Fig.~\ref{fig:r-dist}. A more detailed comparison is included in Appendix~\ref{app:comparison}. Note that only the value of $r$ is physical, while $A$ depends on the specific $t_\mathrm{bin}$ setup. We will solely refer to  $r$ for the simulations, keeping in mind that it is translated from Eq.~\eqref{eq:r-A-relation}.

\begin{figure}
	\includegraphics[width=0.5\textwidth]{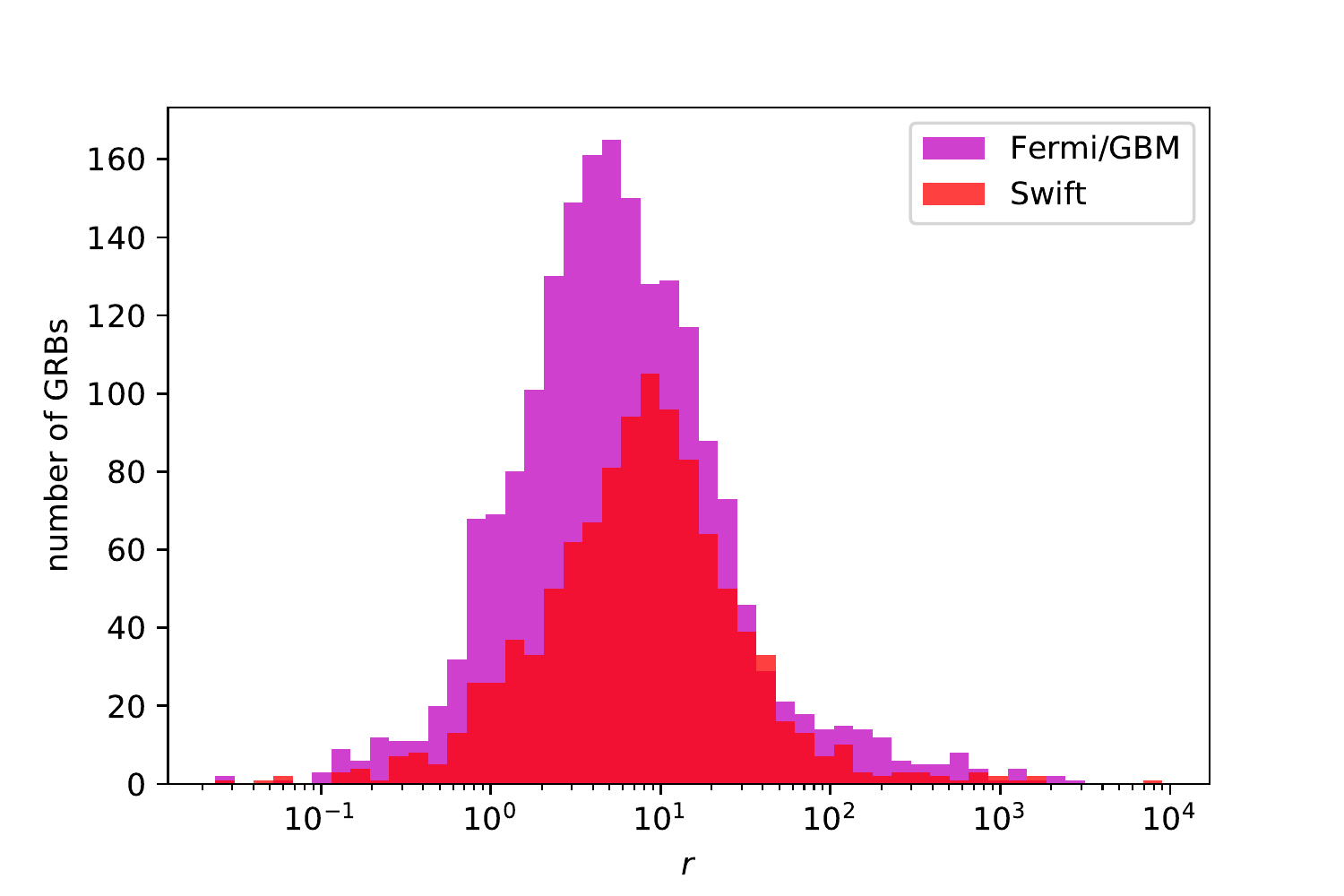}
	\caption{Power ratio $r$ distribution in Fermi/GBM and Swift/BAT. The number of GRBs with $r \geq 1000, 1500, 2000, 3000, 5000, 10000$ is   $10, 4, 3, 1, 1, 0$ for GBM and $5, 3, 1, 1, 1, 0$ for BAT (using 1922 and 1060 GRBs from each catalog which have adequate data to calculate the power spectrum in this frequency range).}\label{fig:r-dist}
\end{figure}

With the scaling and noise parameters determined by observations, we proceed to test the detection algorithm. Here we choose $\mathrm{MVT}=100\mu\mathrm{s}$, which will allow us to probe the interesting mass range $1M_\odot \lesssim M_L \lesssim 100M_\odot$. Larger MVTs could also be considered if needed. But the main effects of the instrumental sensitivity will be similar for any choice of MVT. The light curve simulation is performed on a $[0,1]\,\mathrm s$ interval with  $t_\mathrm{bin}=50\,\mu\mathrm s$ binning.

\begin{figure*}
  \includegraphics[width=0.95\textwidth]{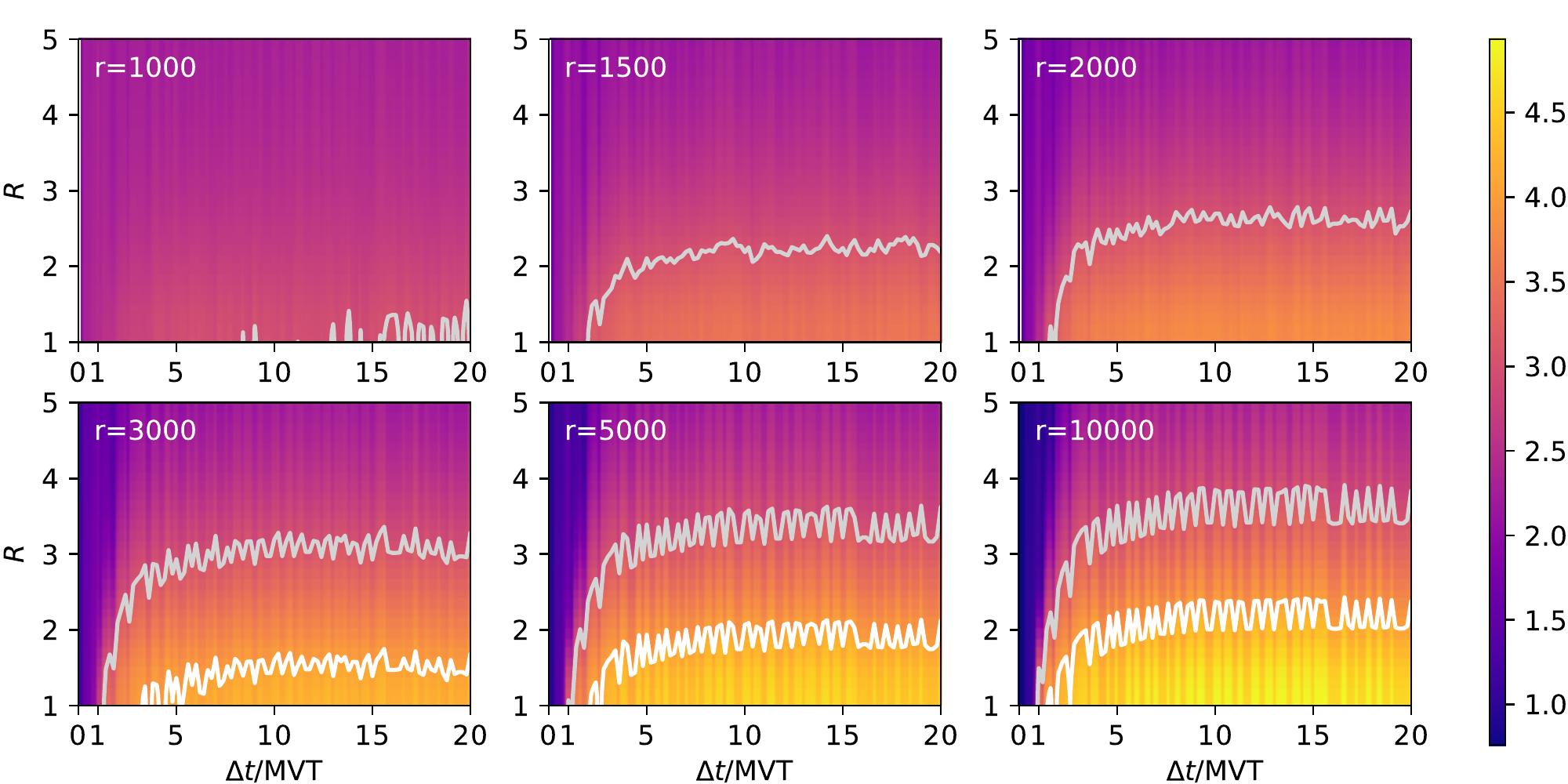}
  \caption{Detection significance (in sigmas) for different time delays $\Delta t$ and magnification ratios $R$ in a simulated lensing experiment. We set $\text{MVT}=1\,\mathrm{ms}$ and use the MVT as the time delay unit. White (gray) lines represent a $3\sigma$ ($4\sigma$) threshold. It can be seen that a higher signal strength $r$ yields a better detection threshold $\overline R$, as expected. The plot shows that $r\sim 5000$ will result in a $3\sigma$ ($4\sigma$) detection at $\overline R=3.0$ ($2.0$). Also, with this $r$, all time delays longer than $\overline{\Delta t}=\mathrm{MVT}$ are detectable.}\label{fig:response}
\end{figure*}

The results are shown in Fig.~\ref{fig:response}. We see that a 3$\sigma$ (4$\sigma$) detection can be achieved with Fermi/GBM or Swift/BAT at $\overline R\!=\!3.0$ ($2.0$), with MVT=$1\,\mu\mathrm{ms}$ and a light curve power ratio $r\!\sim\! 5000$. This ratio is much higher than the typical value for current data. This means that in order to use this method to detect lensed GRB, the noise amplitude (i.e.\ background photon counts) in a future GRB observatory would have to be decreased by a factor $\sqrt{5000/10}\sim 22$.

\section{Discussion}
\label{sec:discussion}

Evidently, a major restriction on the detectability of strongly lensed GRBs is the  background photon noise. Typically, GRB observatories record photons with energy ranging from several keV to several hundreds of keV. In this range, photons from the cosmic X-ray background (CXB) and cosmic gamma-ray background (CGB) will act as noise, with CXB expected to be the dominant source. The rate of noise photons detected in Swift/BAT can be estimated by
\begin{align}
\begin{split}
		\mathrm{rate}=&\left(\frac{\int_{15\mathrm{keV}}^{300\mathrm{keV}}S_\mathrm{CXB}(E)d\ln E}{2.33\ \mathrm{events}\cdot\mathrm{s}^{-1}\cdot\mathrm{cm}^{-2}\cdot\mathrm{sr}^{-1}}\right)\\
	&\times \left(\frac{\mathrm{aperture}}{5200\ \mathrm{cm}^2}\right)\left(\frac{\mathrm{FOV}}{1.4\ \mathrm{sr}}\right) 1.7\times 10^4\ \mathrm{events/sec},
\end{split}
\end{align}
where $S_\mathrm{CXB}(E)$ is the power spectrum of CXB, and $d\ln E$ counts the photon number. The integration is normalized to the value from Ref.~\cite{Ajello:2008xb}, while the aperture size and the BAT field of view (FOV) are normalized as in Ref.~\cite{Gehrels:2004aa}.
Our estimate is consistent with that of Ref.~\cite{Gehrels:2004aa}. It indicates that if one can effectively mask (background) photons originating from outside a region $\lesssim1^{\circ}$, based on the localization of the GRB, the signal-to-noise could be improved enough to enable the detection of strong lensing of GRBs by MACHOs in the $M_L \gtrsim 10\, M_\odot$ range. Instruments using a coded-aperture mask may be able to achieve this (see discussion of BAT in the Appendix).

Another issue worth noting is that only a small fraction of GRBs have the required $\mathrm{MVT}\sim 1\mathrm{ms}$ to probe masses $\gtrsim10\,M_\odot$. A distribution of the MVTs of Fermi/GBM GRBs inferred from the data using a wavelet analysis can be found in Ref.~\cite{MacLachlan:2012cd}. Apparently, variability of a few milliseconds is not uncommon,
and GRBs with suitable MVT can be used to construct a valid dataset.

\section{Conclusions}
\label{sec:conclusions}

In this paper we have investigated the use of GRBs to seek evidence of MACHO dark matter in the mass range $M_L \gtrsim 10\, M_\odot$ via strong gravitational lensing. We have shown that performing an autocorrelation test for $\mathcal{O}(1000)$ GRBs with appropriate minimum variability timescales and background noise amplitudes can potentially result in sensitivity to $f_{\textrm{DM}}\sim1\%$. This sensitivity in the mass range $10-100\, M_\odot$ of interest requires GRBs with MVT $\sim1$ ms.

The strength of the limits on $f_\mathrm{DM}$ depends on $\overline R$, the maximum detectable magnification (see Fig.~\ref{fig:constraints}).
Fermi/GBM or Swift/BAT, however, are not capable of doing so. To achieve the level of $\overline R$ mentioned in Fig.~\ref{fig:constraints}, we would need a signal-to-noise power ratio of $r\sim 5000$.
The distributions of power ratios for these two observatories are centered at $r\sim 10$ (with BAT improving on GBM by a factor $\sim3$). In order for a future GRB observatory to fulfill the constraining potential of $f_{\rm DM}\lesssim1\%$,  the noise amplitude (i.e.\ photon count) would have to be decreased further by an order of mangitude. This can potentially be done with improved localization accuracy.

\begin{acknowledgements}
	We are grateful to Andrew Fruchter and Brice M\'enard for very valuable conversations;  Bo Zhang and Bing Zhang for instructive information on the simulation of GRB light curves (we especially appreciate receiving a tailored set of light curve simulations  from Bo Zhang which we hope to investigate together in future work); Valerie Connaughton for a detailed guide on accessing the Fermi/GBM database; and  Brad Cenko and Amy Lien for help with the Swift/BAT database and software. This work was supported by NSF Grant No. 0244990, NASA  NNX17AK38G, and the Simons Foundation.
\end{acknowledgements}

\appendix

\section{Comparison between Fermi/GBM and Swift/BAT}
\label{app:comparison}

\begin{figure*}[h!]
	\includegraphics[width=0.9\textwidth]{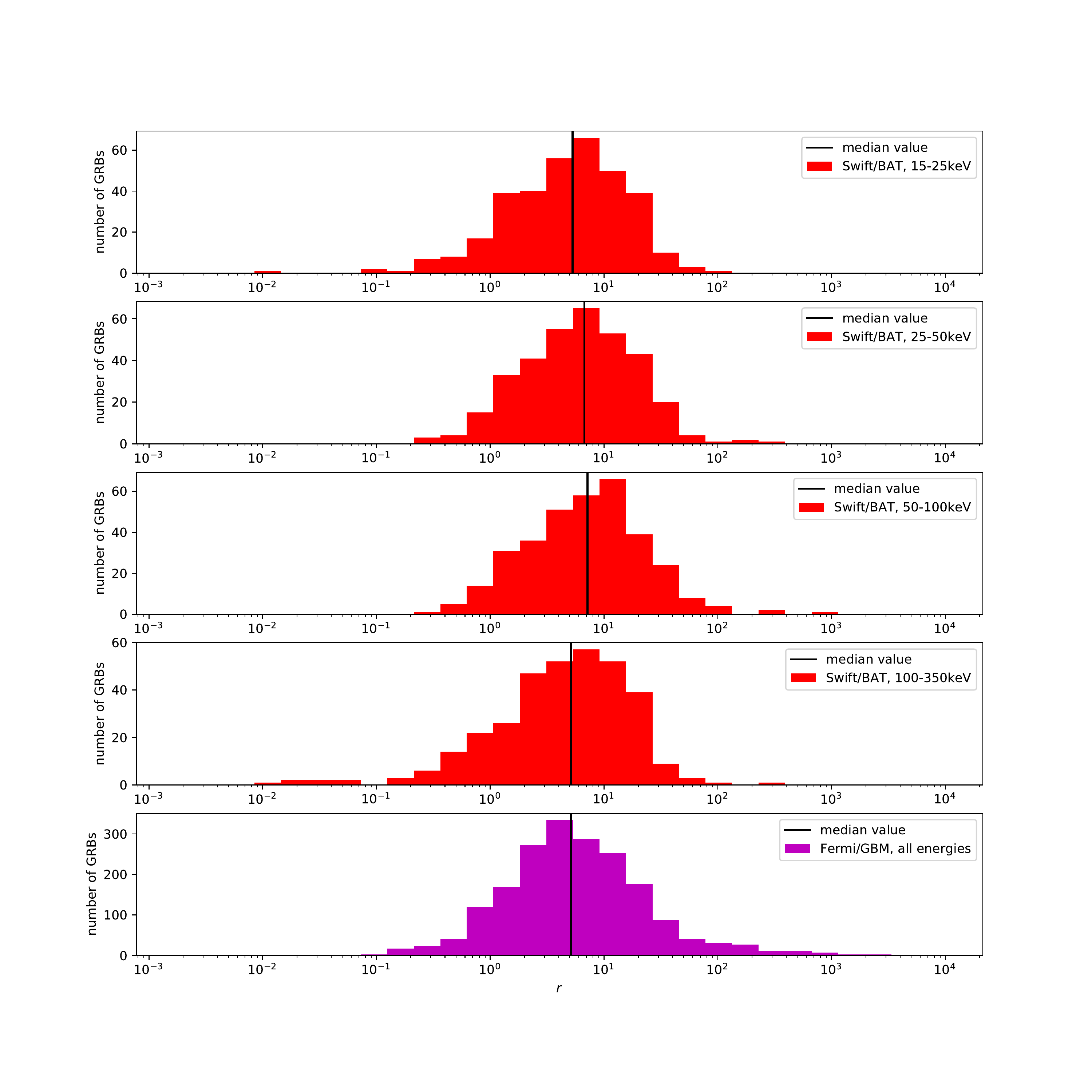}
	\vspace{-0.5in}
	\caption{Power ratio $r$ distribution in Fermi/GBM and Swift/BAT, with BAT power ratio shown for each energy bin. Black vertical lines are the median values of each distribution. In the relevant energy bins ($E<100\mathrm{keV}$), BAT results can be one order of magnitude better than GBM results.}\label{fig:r-dist-ebins}
\end{figure*}

One major difference between Fermi/GBM and Swift/BAT is that BAT uses a coded-aperture mask, which gives it a better imaging (i.e.\ localization) ability than GBM. This advantage is especially useful at lower photon energies: the CXB is stronger at lower energy, so BAT will significantly reduce more noise photons than GBM due to the coded-aperture mask. A comparison in Fig.~\ref{fig:r-dist-ebins} shows that this improvement is as much as one order of magnitude in terms of the power ratio $r$. Unfortunately, this is still not enough for our purpose (another order of magnitude improvement would be required).

\end{document}